%% file: main.tex
\documentclass[conference,10pt,letterpaper]{IEEEtran}
\usepackage{cite}
\usepackage{amsmath,amssymb,amsfonts}
\usepackage[bookmarks=true,breaklinks=true,letterpaper=true,colorlinks,citecolor=blue,linkcolor=blue,urlcolor=blue]{hyperref}
\usepackage{algorithmic}
\usepackage{graphicx}
\usepackage{textcomp}
\usepackage{xcolor}
\def\BibTeX{{\rm B\kern-.05em{\sc i\kern-.025em b}\kern-.08em
    T\kern-.1667em\lower.7ex\hbox{E}\kern-.125emX}}
\usepackage{multirow}
\usepackage{pifont}

\usepackage{algorithmic}
\usepackage{graphicx}
\usepackage{textcomp}
\usepackage{xcolor}
\usepackage{xcolor,colortbl}
\usepackage{fancyhdr}
\usepackage{multirow}
\usepackage{enumitem}
\usepackage{nccmath}
\usepackage{tabularx}
\usepackage{adjustbox}
\usepackage{wrapfig}
\usepackage{booktabs}

\usepackage{hyperref}

\usepackage{enumitem}

\usepackage{amsmath}
\usepackage{cleveref}

\crefformat{section}{\S#2#1#3} % see manual of cleveref, section 8.2.1
\crefformat{subsection}{\S#2#1#3}
\crefformat{subsubsection}{\S#2#1#3}

\newcommand{\name}{framework}

\definecolor{darkgreen}{rgb}{0.0, 0.5, 0.0}
\definecolor{alizarin}{rgb}{0.82, 0.1, 0.26}
\newcommand{\A}{\textcolor{darkgreen}{available}}
\newcommand{\NA}{\textcolor{alizarin}{no alternative yet}}

\begin{document}

\title{\fontsize{12}{12}\selectfont{
\textbf{Modeling PFAS in Semiconductor Manufacturing to Quantify Trade-offs \\ in Energy Efficiency and Environmental Impact of Computing Systems}}}
% \textbf{PFASware: Quantifying the Environmental Impact of Per- and Polyfluoroalkyl \\ Substances (PFAS) in Computing Systems}}}
%PFASeptive Systems

% % \author{\fontsize{11}{11}\selectfont
% \author{\fontsize{10}{10}\selectfont
% % Confidential Draft --- Do NOT Distribute!
% Mariam Elgamal\textsuperscript{$\dagger$},
% Abdulrahman Mahmoud, %\textsuperscript{1}
% Gu-Yeon Wei,
% David Brooks
% and Gage Hills\\

% \fontsize{10}{12}\selectfont 
% \fontsize{8}{8}\selectfont 
% Harvard School of Engineering and Applied Sciences (SEAS)
% \hspace{1em} \textsuperscript{$\dagger$}email: mariamelgamal@g.harvard.edu
% \vspace{-5ex}
% }

\author{\fontsize{10}{10}\selectfont
Mariam Elgamal\textsuperscript{1},
Abdulrahman Mahmoud\textsuperscript{2}, %\textsuperscript{1}
Gu-Yeon Wei\textsuperscript{1},
David Brooks\textsuperscript{1},
and Gage Hills\textsuperscript{1}\\

% \fontsize{10}{12}\selectfont 
\fontsize{8}{8}\selectfont 
\textsuperscript{1}Harvard School of Engineering and Applied Sciences (SEAS), \textsuperscript{2}Mohamed Bin Zayed University of AI (MBZUAI) \\
% \hspace{1em} \{mariamelgamal, guyeon, dbrooks, ghills\} @g.harvard.edu \\
\hspace{1em} mariamelgamal@g.harvard.edu, abdulrahman.mahmoud@mbzuai.ac.ae, \{guyeon, dbrooks, ghills\} @seas.harvard.edu \\
\vspace{-3em}
}

\maketitle
\input{text/abstract}
\input{text/intro}
\input{text/background}
\input{text/framework}
\input{text/methodology}
\input{text/results}

\input{text/opportunities}
\input{text/conclusion}

% \section*{Acknowledgments}
% We acknowledge valuable support from \todo{Salata and NSF Expeditions in Computing}. NSF Carbon Connect? We thank Carole-Jean Wu, Heidi Pickard, Nestor Cuevas, and Georgios Kyriazidis for their valuable discussions.  
% \end{acks}

\clearpage
% INSERT REFERENCE LIST
% \bibliography{references}
\bibliography{references}

\end{document}

%% file: text/abstract.tex
\noindent \textbf{\textit{Abstract}}---The electronics and semiconductor industry is a prominent consumer of per- and poly-fluoroalkyl substances (PFAS), also known as forever chemicals. PFAS are persistent in the environment and can bioaccumulate to ecological and human toxic levels. 
Computer designers have an opportunity to reduce the use of PFAS in semiconductors and electronics manufacturing, including integrated circuits (IC), batteries, displays, etc., which currently account for a staggering 10\% of the total PFAS fluoropolymers usage in Europe alone.
In this paper, we present a \name~where we \textbf{(1)} quantify the environmental impact of PFAS in computing systems manufacturing with granular consideration of the metal layer stack and patterning complexities in IC manufacturing at the design phase, \textbf{(2)} identify contending trends between embodied carbon (carbon footprint due to hardware manufacturing) versus PFAS. For example, manufacturing an IC at a 7 nm technology node using EUV lithography uses 18\% less PFAS-containing layers, compared to manufacturing the same IC at a 7 nm technology node using DUV immersion lithography (instead of EUV) unlike embodied carbon trends, and \textbf{(3)} conduct case studies to illustrate how to optimize and trade-off designs with lower PFAS, while meeting power-performance-area constraints. We show that optimizing designs to use less back-end-of-line (BEOL) metal stack layers can save 1.7$\times$ PFAS-containing layers in systolic arrays.

% We show that manufacturing an IC with a 16 nm technology node results in 15\% less PFAS volume than manufacturing with a 28 nm lagging technology node due to area savings. We also show that manufacturing an IC at a 7 nm technology node using Extreme Ultraviolet (EUV) lithography uses 20\% less volume of PFAS-containing chemicals, compared to manufacturing the same IC at a 7 nm technology node using Deep Ultraviolet (DUV) immersion lithography (instead of EUV).

%% file: text/intro.tex
\vspace{-0.3em}
\section{Introduction}
\label{sec:intro}
\vspace{-0.3em}
The environmental impacts of computing systems go beyond their carbon footprint and water consumption. The chemicals and materials used in the semiconductor and electronics manufacturing processes have environmental and human health impacts that require our immediate attention as computer designers and engineers. A class known as per- and poly-fluoroalkyl substances (PFAS)---also referred to as forever chemicals---constitutes more than 16,000 chemicals used in manufacturing across global industries~\cite{PFAsNature2023article}. PFAS are a physiochemically diverse class of synthetic chemicals, 
%ranging from salts to polymers to acids, related due to their use of chemically stable perfluorinated carbon chains. PFAS 
containing one or more fully fluorinated methyl (three carbon-fluorine bonds) or ethylene (two carbon-fluorine bonds) carbon atoms~\cite{OECD2021PFAS}. Due to their bioaccumulation, human toxicity, and environmental impact, PFAS have been under extensive public, scientific and regulatory action globally~\cite{3MPFAS2025exit, EPA2022Advisory, ECHA2023PFASRestrictionReport}. Across the electronics and semiconductor industry, PFAS are used in a wide variety of capacities including manufacturing computing ICs, displays, batteries, cooling liquids for thermal management in datacenters, and more~\cite{PFAsNature2023article, Apple2022whitepaper, chemsec}. With the rising proliferation of electronics and computing chips, PFAS use in electronics are expected to grow 10\% annually, primarily driven by semiconductor manufacturing and production~\cite{ECHA2023PFASRestrictionReport}.

\begin{figure}[t!]
\centering
\includegraphics[width=\columnwidth]{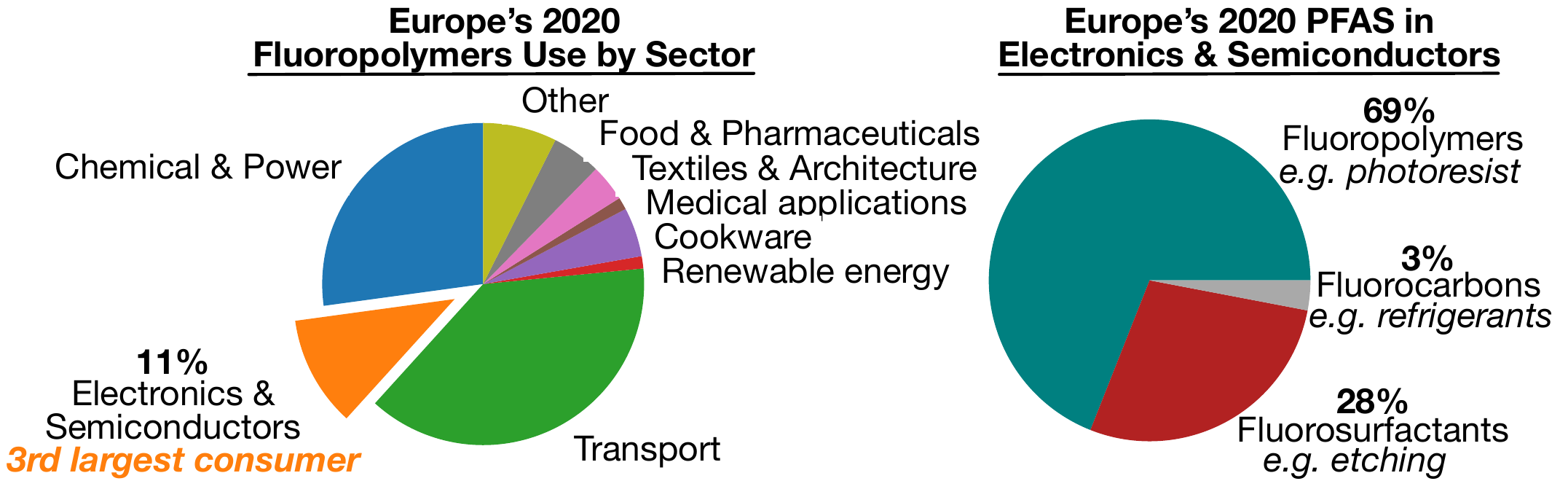}
\vspace{-0.7cm}
\caption{Electronic and semiconductor industries are the third largest consumer of fluoropolymers across sectors in Europe in 2020 (left)~\cite{fluoropolymer2020EUelectronics}. In Europe alone, over 4.21 kilotons of PFAS were used in semiconductors and electronics manufacturing (right)~\cite{PFAsNature2023article}. Computer designers and researchers need to identify and incorporate design optimizations to reduce PFAS at the design phase.}
%% Gage comment: Make sure both of theses figures show up on the first page. I think you can trim the abstract. The abstract can likely be very short, maybe 3-5 sentences.
% \vspace{-1.1em}
\label{fig:PFAS-pie}
\vspace{-1.6em}
\end{figure}
% \begin{figure}[t!]
% \centering
% \includegraphics[width=\columnwidth]{fig/PFAS_DATE2025_lifecycle_v1.pdf}
% \vspace{-2.2em}
% \caption{There are three possible contaminant streams for PFAS in semiconductors and electronics: most commonly \textbf{atmospheric} and \textbf{aqueous} releases throughout manufacturing, and \textbf{leachate} release through the soil due to e-waste that contaminates the environment in the long-term.} %and ground water on the
% \label{fig:PFAS-diagram}
% \vspace{-2em}
% \end{figure}

\input{tables/tab_PFAS_alt}
As more fabrication facilities are built globally, the increasing use of PFAS during manufacturing as well as PFAS contaminants, whether through wastewater, emissions, or increasing e-waste~\cite{KUMAR2017ewaste}, are pressing environmental issues for the semiconductor and computing industry at large. In this paper, we take a data-driven approach to studying PFAS in ICs manufacturing. We model PFAS use in different lithography manufacturing steps and identify design optimization trade-offs between PFAS, embodied carbon, power and performance. To our knowledge, this is the \textbf{first work to enable researchers and designers to model and quantify PFAS from hardware manufacturing while proposing design strategies to minimize PFAS at the design phase.} Our key contributions:
\vspace{-0.2em}
\begin{enumerate} [wide, labelindent=8pt]
    \item[\scalebox{1.05}{\ding{172}}] We propose an analytical modeling framework to quantify the amount of PFAS used in logic semiconductor manufacturing (\cref{sec:framework}). Our framework leverages detailed fabrication facility characterization and existing literature on semiconductor photolithography and processing complexity, to enable designers to estimate the PFAS consumption of their designs during manufacturing at the design phase.
    % \item[\scalebox{1.05}{\ding{173}}] We propose a PFAS-efficiency optimization metric based on manufacturing, design and environmental factors (\cref{sec:metric}). 
    \item[\scalebox{1.05}{\ding{173}}] To enable more holistic sustainable computing systems design, we integrate an architectural carbon modeling tool~\cite{ACTGupta2022} in our framework to quantify and trade-off PFAS with carbon emissions due to hardware manufacturing (\cref{sec:methodology} and \cref{sec:PFAScarbon}). 
    \item[\scalebox{1.05}{\ding{174}}] We conduct PFAS-aware design case studies (\cref{sec:results}) and propose design knobs for designers to make less PFAS-containing designs. We demonstrate designing hardware systems with less number of back-end-of-line (BEOL) metal layers can lead to 1.7$\times$ less PFAS in manufacturing (\cref{sec:metalstack}). 
    % \item[\scalebox{1.05}{\ding{175}}] To catalyze further investigation into quantifying PFAS in semiconductor manufacturing, we provide a \textbf{public and open-source modeling PFAS framework} to the community.\footnote{The GitHub repository will be made available upon publication.}
\end{enumerate}

% \input{tables/tab_PFAS_alt}
% With the rise of chiplet design and a growing metal stack with more advanced technology nodes, the environmental sustainability of computing is a major question. 
% Prior works have characterized and modeled the carbon footprint of computing systems, including emerging technologies~\cite{chasingGupta2021, ACTGupta2022, imec2020}. More recently, PFAS have become .

% However, environmental sustainability extends beyond carbon footprint and water consumption. 
%% Highlight results here
% \fixme{We show that manufacturing an IC at a 16 nm technology node results in 15\% less PFAS volume than manufacturing the same IC at a lagging 28 nm technology node due to area savings. We also show that manufacturing an IC at a 7 nm technology node with EUV described in~\cite{imec2020,TSMCn7}, which comprises four EUV-patterned metal layers, uses 20\% less volume of PFAS-containing chemicals, compared to manufacturing a 7 nm technology node with DUV immersion lithography (193 nm wavelength) to pattern these same four layers as describe in~\cite{imec2020}. Additionally, the relative power is 15\% lower (better) and the relative performance is 10\% higher (better) for a 7 nm process technology node manufactured with EUV at Taiwan Semiconductor Manufacturing Company (TSMC)~\cite{TSMCn3}.}

%% file: tables/tab_PFAS_alt.tex
\begin{table*}[t!]
\centering
% \vspace{-1em}
\caption{A non-exhaustive summary of PFAS in electronics and semiconductors, their use, function, and availability of alternative substitutes or lack thereof. Integrated circuits manufacturing lack the most in PFAS-free alternatives.}
%~\cite{ChemSec2023CheckYourTech}
\label{tab:pfas-catalog}
\vspace{-0.7em}
\scalebox{0.87}{
\begin{tabular}{c|c|c|c|c}
\toprule
\textbf{Part}   &  \textbf{PFAS Type} & \textbf{Use} & \textbf{Function} & \textbf{Alternatives} \\
\hline \hline
& photopolymer & photolithography & photoresist & \NA~(research stage)~\cite{commissioned2022PFASlitho, SIAPFAS2023photolitho} \\
& photopolymer & photolithography & photoacid generator & \NA~(research stage)~\cite{commissioned2022PFASlitho, SIAPFAS2023photolitho} \\
& short fluoropolymers & anti-reflective coating & low refractive index & \A~but not demonstrated in DUV~\cite{commissioned2022PFASlitho}\\
Integrated Circuits & short-chain PFAS & developers & remove unwanted resist pattern & \NA~(research stage)~\cite{commissioned2022PFASlitho} \\
& PFAS additives & rinsing solutions & low-surface tension & \NA~(research stage)~\cite{SIAPFAS2023wetchemistry} \\
& fluorocarbon gases & dry etching & precision in etching & \NA~(research stage)~\cite{PFAsNature2023article} \\
& fluorosurfactants~\cite{3MElectronicSurfactant4200} & wet etching & improve coating quality & \A~(testing \& trials stage)~\cite{PFASfree2023surfactants, SIAPFAS2023wetchemistry} \\
%& PFPEs & vapor phase soldering & heat transfer medium & \NA \\
& fluoropolymers & spin-on dielectrics & leakage blocker & \A~\cite{dupont-spinon} \\
\hline
Datacenters & fluorocarbons~\cite{datacenterPFAScooling2023} & cooling liquids & refrigerants; thermal management & \A~(research stage)~\cite{ChemSec2023CheckYourTech}\\
\hline
% Photonics & & & & \\
% \hline
PCBs & fluoropolymer~\cite{PCBTeflon} & laminate material & flame retardant; dielectric & redesign equipment \& product dimensions~\cite{SIAPFAS2023packaging}\\
& fluoropolymer & protective coating & temperature stable; dust repellent & \textcolor{darkgreen}{multiple} \A~\cite{PCBcoating_PFASfree} \\
\hline
Capacitors & fluoropolymers & dielectric films & dielectric strength& \textcolor{darkgreen}{multiple} \A~\cite{ChemSec2023CheckYourTech} \\
% & & liquid impregnates & thermal resistance & \textcolor{darkgreen}{multiple} \A \\
\hline 
Acoustic equipment & fluoropolymers & piezoelectric materials & mold into thin, flexible sheets & \A~depending on product function\\
& fluoropolymers & vent membranes & hydrophobic & \NA~(research stage) \\
\hline
Displays & fluorinated compounds & LCD & dipole moment & \A~e.g., LED or plasma screens~\cite{ChemSec2023CheckYourTech}\\ %other display technologies such as 
% & fluoropolymers & flat panel & dust repellent; resist static electricity & \\
\hline
Wiring \& cables & fluoropolymers & insulating layer & corrosion, thermal, cracking resistant & \A~depending on needed function\\
\hline
Lithium-ion batteries & fluoropolymers & binder & electrochemical stability & \NA~(research stage) \\
& PFAS salts \& additives~\cite{lithiumionbattery2023PFAS} & electrolyte & increased performance \& durability & \A \\
% \hline
% Coatings & & & & \\
\bottomrule
\end{tabular}
}
\vspace{-1.8em}
\end{table*}

%% file: text/background.tex
% \input{tables/tab_PFAS_alt}
\vspace{-0.7em}
\section{Background}
\vspace{-0.3em}
In Europe alone in 2020, European Chemicals Agency (ECHA) estimates the amount of PFAS used in electronics and semiconductor manufacturing to reach 4.21 kilotonnes~\cite{PFAsNature2023article}. Figure~\ref{fig:PFAS-pie} (right) shows that approximately 69\% of those PFAS  come from fluoropolymers, 28\% are fluorosurfactants used for surface wetting and modification, and 3\% are fluorocarbons which are small molecule liquids and gases, some of which are greenhouse gases used in thermal management coolants. Another analysis shows that the electronics and semiconductors industry is the \textit{third largest consumer of fluoropolymers} after the transportation and the chemical and power sectors as shown in Figure~\ref{fig:PFAS-pie} (left), amounting to 11\% of total fluoropolymers sold in the EU in 2020~\cite{fluoropolymer2020EUelectronics, PFAS2024DATEtoappear}. 

\begin{figure}[t!]
\centering
\includegraphics[width=\columnwidth]{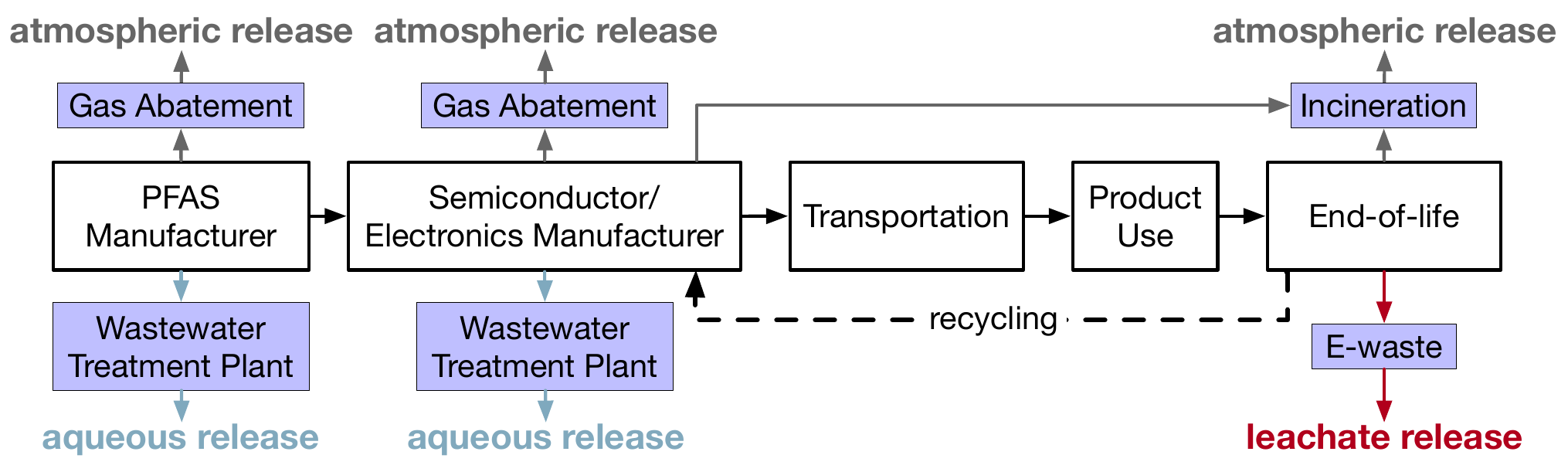}
\vspace{-2.4em}
\caption{There are three possible contaminant streams for PFAS in semiconductors and electronics: most commonly \textbf{atmospheric} and \textbf{aqueous} releases throughout manufacturing, and \textbf{leachate} release through the soil due to e-waste that contaminates the environment in the long-term.} %and ground water on the
\label{fig:PFAS-diagram}
\vspace{-1.8em}
\end{figure}

Many PFAS are environmentally persistent and bioaccumulative, and have been found in water, soil, and air (including the arctic)~\cite{PFAS2022Science, EPA2022Advisory}. 
% Several PFAS have been linked to major human health risks, including cancer, liver damage, immune system suppression and more~\cite{PFOA2022Cancer, PFAS2020toxicityReview}. 
While the toxicological data of most PFAS are currently largely undetermined~\cite{PFOA2022Cancer, PFAS2020toxicityReview}, the potential health consequences and bioaccumulation of forever chemicals stipulate the alarming need for minimizing the use of PFAS whenever possible. 
% In 2022, Apple was one of the first technology companies to release a white paper on their commitment to phase out PFAS from their products~\cite{Apple2022whitepaper}. 
Only few technology companies, such as Apple~\cite{Apple2022whitepaper}, have announced the phasing out of PFAS in their products. While PFAS are mostly safe during product use, many safety concerns arise throughout the manufacturing supply chain and disposal of computing systems. Throughout the computing system's lifecycle, there are three possible flows for PFAS to be released into the environment, namely atmospheric release through air, aqueous release through water, and leachate release through soil. Figure~\ref{fig:PFAS-diagram} illustrates these three potential PFAS contamination streams for semiconductors.
% throughout a computing system lifecycle.

% While 
PFAS remediation methods, i.e. removing PFAS contaminants from water sources and soil, are important solutions towards limiting human exposure to \textit{already existing PFAS}. However, they are insufficient to overcome PFAS pollution across industries~\cite{ChemSec2023PFACleanup,c&en2024PFASdestruction}. Recent studies show that wastewater treatment plants do not fully remove or eradicate PFAS from semiconductor fabrication facilities' wastewater. Furthermore, certain PFAS are systematically resistant to current wastewater treatments~\cite{PFAS2024Taiwan, PFAS2021ElectronicsWastewater}. \textit{This highlights the importance of reducing PFAS-containing chemicals in manufacturing, and even prior at the design phase.} Therefore, there is a critical need to find effective PFAS-free alternatives and minimize PFAS when their use is necessary in computing (e.g. photolithography). In Table~\ref{tab:pfas-catalog}, we show different types of PFAS used in a variety of electronics and semiconductor manufacturing processes, as well as the current availability of viable PFAS-free alternatives, or lack thereof.
% \input{tables/tab_PFAS_alt}

% \fixme{SIA survey paper..}
% Based on the SIA manufacturing
\begin{wrapfigure}{rt}{0.43\columnwidth}
  \begin{center}
  \vspace{-2em}
    \includegraphics[width=0.37\columnwidth]{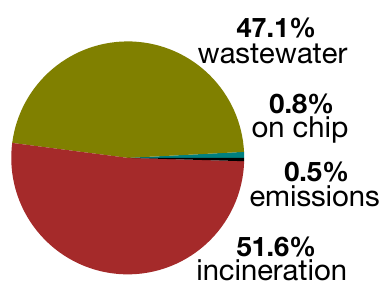}
  \end{center}
  \vspace{-1.5em}
  \caption{Majority of PFAS used in semiconductor manufacturing are either incinerated or go to wastewater~\cite{SIA2024PFASpaper}, signifying major opportunities to reduce PFAS.}
  \vspace{-1em}
  \label{fig:PFAS-disposal}
\end{wrapfigure}
% As computer designers, we have an opportunity to reduce PFAS usage in computing systems at the design phase.

The primary use of PFAS in semiconductor manufacturing is in photolithography, where only 0.8\% of PFAS used remain on chip~\cite{SIA2024PFASpaper}, and the majority are either disposed of in wastewater or incinerated (Figure~\ref{fig:PFAS-disposal}). Building environmentally sustainable computing systems, including low PFAS consumption, requires collaborations across the computing stack, from design to manufacturing to disposal, and collective collaboration among academia and industry. 
Researchers and designers across the computing stack have an opportunity to identify trade-offs and incorporate optimizations for lower environmental impacts of PFAS at the design phase, especially due to the current lack of PFAS-free alternatives in photolithography and ICs manufacturing. PFAS are primarily used in:

\begin{figure*}[t!]
\centering
\includegraphics[width=0.95\textwidth]{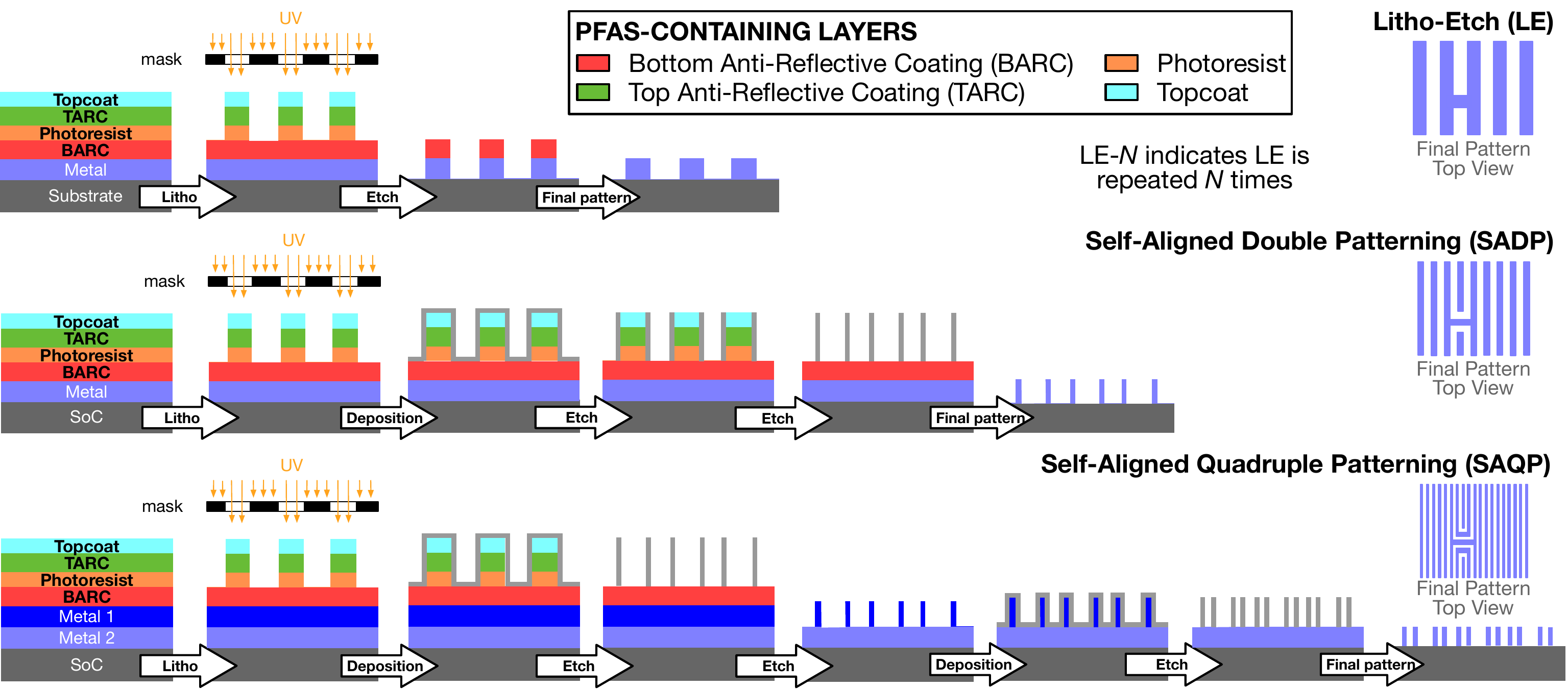}
\vspace{-0.3cm}
\caption{Lithography is a key fundamental technology for semiconductor manufacturing. A PFAS-containing layer in lithography includes chemicals, such as bottom and top antireflective coats, photoresist and topcoat. We illustrate example process complexities for immersion 193 nm DUV (ArFi) and EUV industry-wide lithography options such as LE-$N$, SADP and SAQP. $N$ indicates number of litho-etch sequences, ranging from single to quadruple patterning.}
\vspace{-1.6em}
\label{fig:PFAS-granularity-methodology}
% \vspace{-1.1em}
\end{figure*}

% Lithography is one of the key fundamental technologies for semiconductor manufacturing, and is also one of the primary consumers of PFAS-containing chemicals~\cite{ECHA2023PFASRestrictionReport,plummer2023integrated}.
% In semiconductor photolithography, 
% Lithography is a key fundamental technology for semiconductor manufacturing, where 
% In IC manufacturing, PFAS are primarily used in:
\vspace{-0.3em}
\begin{enumerate}[wide, labelindent=8pt]
    \item \textbf{Photoresists.} These are photopolymers used to pattern micro and nano structures on a substrate by changing its solubility upon exposure to short wavelength radiation, such as extreme ultraviolet (EUV) or deep ultraviolet (DUV). This process requires yield above 99\% given the many process steps needed to manufacture advanced semiconductor chips~\cite{commissioned2022PFASlitho}.

    \item \textbf{Antireflective coatings (ARCs).} Used for their low refractive index, to prevent light interference reflected from substrate, and as a barrier layer. 
    These include top and bottom antireflective coats (TARCs and BARCs, respectively).  
    
    \item \textbf{Other coatings.} Such as topcoats are used to prevent leaching of photoactive components and protect photoresist from immersion water used in 193 nm immersion DUV~\cite{commissioned2022PFASlitho}. Furthermore, dielectric spin-on coatings improve coating uniformity in photoresists and ARCs and prevent defects.
    % \item \textbf{Nanoimprint lithography fluoropolymers.} Nanoimprint lithography is a future lithography option and is still being developed with PFAS-containing chemicals. 
\end{enumerate}
% In this paper we leave nanoimprint lithography for future work.
We identify lithography steps which include PFAS chemicals and refer to them as PFAS-containing layers in this work. \cref{sec:framework} details the design and implementation of the proposed PFAS modeling tool for semiconductor manufacturing. 

%% file: text/framework.tex
\vspace{-0.5em}
\section{Framework for PFAS-Aware Systems Design}
\label{sec:framework}
\vspace{-0.3em}
Computer designers have a key opportunity to minimize PFAS at the design phase.
In this section, we detail how we quantify and model the amount of PFAS used in semiconductor manufacturing per metal layer of IC design based on fabrication specifications (\cref{sec:model}). We detail the inputs and outputs of our framework (\cref{sec:PFASframework}), and discuss limitations and future steps for researchers and designers to better account for PFAS in computing systems (\cref{sec:limitations}).

\vspace{-0.5em}
\subsection{PFAS and Semiconductor Manufacturing Analytical Model}
\label{sec:model}
\vspace{-0.2em}
The amount of PFAS used in semiconductor manufacturing is highly dependent on complexity of patterning and metal layer stack. 
In Figure~\ref{fig:PFAS-granularity-methodology}, we illustrate the process complexities and PFAS-containing layers of the most commonly used lithography options  in semiconductor manufacturing, including Litho-Etch (LE), Self-Aligned Double Patterning (SADP), and Self-Aligned Quadruple Patterning (SAQP). Depending on the process flow and chemicals used, the amount of PFAS and combined use of antireflective coats, spin-on coats and topcoats with a specific photoresist vary. For example, one process flow may include BARC and photoresist only~\cite{imec2017recipeexample}. 
% The patterning complexity and process flow of a process technology node can be updated and specified in more detail within our model.
Our model offers the flexibility for designers and engineers to update and customize the patterning complexity and process flow of a given process technology node, allowing for the integration of user-defined fabrication specifications.

% The patterning does not only include lithography steps, it also includes deposition, etching and cleaning steps. 
% \item \textbf{Number of processing steps.}

\input{tables/tab_num_masks}

% \begin{figure}[t!]
% \centering
% \includegraphics[width=\columnwidth]{fig/PFAS_DATE2025_metalstack_v3.pdf}
% \vspace{-0.8cm}
% \caption{Number of PFAS-containing lithography layers in chip manufacturing across process technology nodes, based on our \name~model. PFAS used in manufacturing are dependent on the design's BEOL metal stack and patterning complexity. Unlike energy consumption and carbon footprint, PFAS do not exhibit consistent upward trends with more advanced technology nodes.}
% \vspace{-1.8em}
% \label{fig:PFAS-stack}
% % \vspace{-1em}
% \end{figure}

We base our current PFAS model on the metal stacks presented in~\cite{imec2020, TSMCmetalstack}, and approximate the amount of PFAS-containing layers based on the lithography steps which typically use PFAS-containing photoresist, ARCs, topcoats and other coatings. In semiconductor manufacturing, masks are used to lithographically pattern features on silicon wafers. Therefore, we use the number of lithography masks as a proxy to estimate the amount of PFAS used in manufacturing (Table~\ref{tab:num_masks}). 
% i.e. \# PFAS$_\text{litho}$
The analytical PFAS modeling equations are: 
% \vspace{-0.5em}
% \scalebox{0.94}{
% \begin{equation}
% \label{eq:analytical_PFAS}
% \begin{aligned}
% & \textrm{PFAS}_\textrm{wafer} (\text{FEOL, MOL, BEOL, patterning complexity}) \\
% &\propto \text{Photoresist} + \text{TARC} + \text{BARC} + \text{Topcoat} \\
% % \text{Spin-on} +
% & \approx \text{Number of lithography masks}\\
% & = \text{\# PFAS}_\text{litho}
% \end{aligned}
% \end{equation}
% }
% \vspace{-1em}
% \scalebox{0.94}{
% \begin{equation}
% \label{eq:analytical_PFAS_chip}
% \begin{aligned}
% \textrm{PFAS}_\textrm{chip manufacturing} &= \text{\# PFAS}_\text{litho} \times \frac{\text{Area}}{\text{Yield}} 
% \end{aligned}
% \end{equation}
% }
% \vspace{-1.5em}
\vspace{-0.5em}
\begin{equation}
\label{eq:analytical_PFAS}
\begin{aligned}
& \textrm{PFAS}_\textrm{wafer} (\text{FEOL, MOL, BEOL, patterning complexity}) \\
& \propto \text{Photoresist} + \text{TARC} + \text{BARC} + \text{Topcoat} \\
& \approx \text{Number of lithography masks} \\
& = \text{\# PFAS}_\text{litho}
\end{aligned}
\end{equation}
\vspace{-1em}
\begin{equation}
\label{eq:analytical_PFAS_chip}
\textrm{PFAS}_\textrm{chip manufacturing} = \text{\# PFAS}_\text{litho} \times \frac{\text{Area}}{\text{Yield}}
\end{equation}
\vspace{-1.5em}
% \begin{figure}[t!]
% \centering
% \includegraphics[width=\columnwidth]{fig/PFAS_DATE2025_metalstack_v3.pdf}
% \vspace{-0.8cm}
% \caption{Number of PFAS-containing lithography layers in chip manufacturing across process technology nodes, based on our \name~model. PFAS used in manufacturing are dependent on the design's BEOL metal stack and patterning complexity. Unlike energy consumption and carbon footprint, PFAS do not exhibit consistent upward trends with more advanced technology nodes.}
% \vspace{-1.5em}
% \label{fig:PFAS-stack}
% % \vspace{-1em}
% \end{figure}

\begin{figure}[t!]
\centering
\includegraphics[width=\columnwidth]{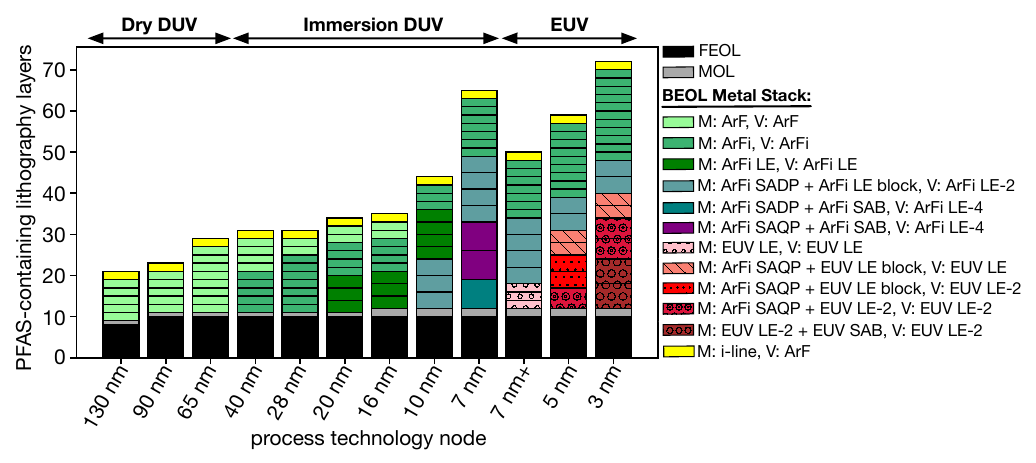}
\vspace{-0.8cm}
\caption{Number of PFAS-containing lithography layers in chip manufacturing across process technology nodes, based on our PFAS model. PFAS used in manufacturing are dependent on the design's BEOL metal stack and patterning complexity. Unlike energy consumption and carbon footprint, PFAS do not exhibit consistent upward trends with more advanced technology nodes.}
\vspace{-1.2em}
\label{fig:PFAS-stack}
% \vspace{-1em}
\end{figure}

We quantify the amount of PFAS in semiconductor manufacturing for 130 nm to 3 nm process technology nodes. In Figure~\ref{fig:PFAS-stack}, we show the PFAS-containing lithography layers in chip manufacturing across front-end-of-line (FEOL), middle-of-line (MOL) and back-end-of-line (BEOL). The amount of PFAS-containing layers increase with increasing number of lithography steps during IC fabrication. Using direct EUV results in less PFAS-containing layers compared to DUV, which correspond to lower number of masks and process steps (refer Figure~\ref{fig:PFAS-stack}). 
% To illustrate, 7 nm technology nodes using immersion 193 nm (DUV) lithography require \textit{additional} PFAS-containing topcoats and an embedded barrier layer (EBL) to prevent immersion water from leaching into the photoresist during multiple patterning~\cite{commissioned2022PFASlitho}, compared to direct 7 nm EUV. 
However, as feature sizes get smaller, the number of masks and lithography steps increase due to the rising complexity of more advanced technology nodes. 
% This results in an increasing trend for PFAS-containing layers for 5 nm technology nodes and beyond.

% \begin{figure}[t!]
% \centering
% \includegraphics[width=\columnwidth]{fig/PFAS_DATE2025_metalstack_v1.pdf}
% \vspace{-0.8cm}
% \caption{Number of PFAS-containing lithography masks in chip manufacturing across process technology nodes, based on our \name~model. PFAS used in manufacturing are dependent on the design's BEOL metal stack and patterning complexity. Unlike energy consumption and carbon footprint, PFAS do not exhibit consistent upward trends with more advanced technology nodes.}
% \vspace{-1.1em}
% \label{fig:PFAS-steps}
% % \vspace{-1em}
% \end{figure}

% \input{tables/tab_num_masks}
\input{tables/tab_parameters}

% \begin{figure}[t!]
% \centering
% \includegraphics[width=\columnwidth]{fig/PFAS_DATE2025_framework_highlevel_v2.pdf}
% \vspace{-0.8cm}
% \caption{Overview of our \name~design framework for computing systems.}
% \vspace{-2em}
% \label{fig:PFAS-framework}
% % \vspace{-1em}
% \end{figure}

\vspace{-0.3em}
\subsection{PFAS Framework}
\label{sec:PFASframework}
\begin{figure}[t!]
\centering
\includegraphics[width=\columnwidth]{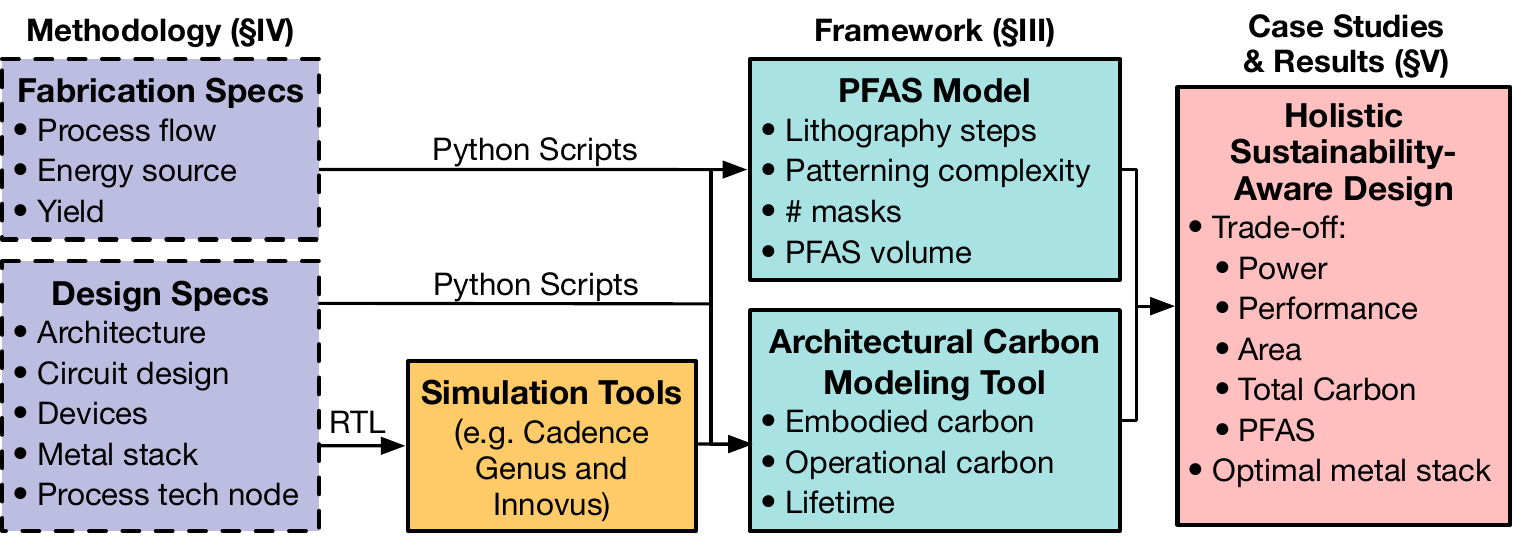}
\vspace{-0.8cm}
\caption{Overview of our PFAS design framework for computing systems.}
\vspace{-1.4em}
\label{fig:PFAS-framework}
% \vspace{-1em}
\end{figure}

\vspace{-0.2em}
We detail our proposed framework for PFAS and carbon aware design in Figure~\ref{fig:PFAS-framework}. Our framework takes input parameters across multiple layers of the computing stack, from fabrication facility specifications, such as yield and process flow, to metal stack specifications and architecture of the design as well as power, performance and die area. Table~\ref{tab:parameters} defines each of the input parameters in our framework. The inputs are fed into our PFAS model and an architectural carbon modeling tool ACT~\cite{ACTGupta2022}. This enables holistic sustainability-aware design by trading-off power, performance, area, PFAS and carbon emissions of different computing systems (refer~\cref{sec:results}). In addition, our model can be extended to include other process flows, metal stacks, and PDKs to quantify and estimate PFAS use for a variety of manufacturing processes at the design phase. For example, we show the estimated number of PFAS-containing layers based on metal line processes of ASAP7 PDK in Table~\ref{tab:ASAP7_num_masks} (table does not include doping masks in FEOL for brevity). 
% We provide more examples in the github repository. 

% Our datasets and results are based on the lithography process steps and metal stacks existing in the literature (e.g. ~\cite{imec2020}). However, 

% \subsection{PFAS Efficiency Metric}
% \label{sec:metric}
\vspace{-0.5em}
\subsection{Limitations}
\label{sec:limitations}
\vspace{-0.2em}
While our PFAS model is the first to enable designers and researchers to quantify PFAS in semiconductor manufacturing at the IC design phase, there are some limitations due to lack of manufacturing data and PFAS quantification transparency across the supply chain.

% \begin{figure}[t!]
% \centering
% \includegraphics[width=\columnwidth]{fig/PFAS_DATE2025_framework_highlevel_v2.pdf}
% \vspace{-0.8cm}
% \caption{Overview of our \name~design framework for computing systems.}
% \vspace{-2em}
% \label{fig:PFAS-framework}
% % \vspace{-1em}
% \end{figure}
% \input{tables/tab_ASAP7_masks}

First, the environmental impact of PFAS in computing is complex and understudied. Quantifying PFAS is a two-step process. First: environmental scientists need to \textit{detect} PFAS, and second: they need to be able to \textit{measure} PFAS. However, unlike carbon footprint, which can be measured and converted in units of carbon dioxide equivalent (CO$_{2}e$), measuring PFAS is more challenging. With the variation in PFAS chain lengths (i.e. number of perfluorinated carbon), PFAS concentration (units of parts-per-trillion or $\mu$g/L) while used in some studies~\cite{EST2024SurfactantsSemiconductor}, may not be a sufficient indicator alone for PFAS environmental impact. Alternatively, referring to amount of PFAS in fluorine basis, such as 
% number of perfluorinated carbon chains
chain length~\cite{Pickard2024Bioaccumulation, SIA2024PFASpaper} or molarity equivalence will enable more accurate comparisons between different PFAS. For example, a PFAS substance with 1 perfluorinated carbon can be referred to as C1, versus another with 8 perfluorinated carbons is referred to as C8.  

\input{tables/tab_ASAP7_masks}

% Second, IC foundries are incredibly secretive about their process flows and recipes, as well as chemicals use. In addition, chemical companies are secretive about the chemical composition of PFAS-containing chemicals such as photoresists and topcoats. 
Second, IC foundries and chemical companies maintain strict secrecy regarding their process flows, chemical usage, and the chemical composition of PFAS-containing chemicals like photoresists and ARCs. Even if fabrication data are available, there is a wide variation in process steps and recipes across foundries, including for the same process technology node. To overcome these challenges, we quantify PFAS through the number of PFAS-containing layers in semiconductor manufacturing. Given the growing semiconductor industry, we emphasize the need for public and transparent reporting of the environmental impacts of computing across the supply chain to enable research and industry efforts towards building more sustainable computing systems. 

%% file: tables/tab_num_masks.tex
% \usepackage{multirow}
\begin{table}[t!]
\centering
\caption{Number of lithography process steps and masks per metal line process in semiconductor manufacturing}
\label{tab:num_masks}
\vspace{-1em}
\scalebox{0.8}{
\begin{tabular}{ll|cccccc|c}
\toprule
\multirow{2}{*}{}  & \multicolumn{1}{c|}{\multirow{2}{*}{\textbf{Process}}} & \multicolumn{6}{c|}{\textbf{\# Steps}~\cite{imec2020}} & \multicolumn{1}{c}{\multirow{2}{*}{\textbf{\# Masks}}} \\
& \multicolumn{1}{c|}{} & \multicolumn{1}{c}{\textbf{DryEtch}} & \multicolumn{1}{c}{\textbf{Litho.}} & \multicolumn{1}{c}{\textbf{Metal.}} & \multicolumn{1}{c}{\textbf{Metr.}} & \multicolumn{1}{c}{\textbf{WetEtch}} & \multicolumn{1}{c|}{\textbf{Dep.}} & \multicolumn{1}{c}{}  \\
\hline \hline
  & ArF LE   & 1   & 3  & 1   & 2  & 3 & 0  & 1  \\
\hline
\multirow{6}{*}{\rotatebox[origin=c]{90}{\textbf{DUV}}} & ArFi LE  & 1   & 3  & 1  & 3 & 3 & 0 & 1 \\
& ArFi LE-2   & 3  & 6  & 1   & 7  & 3  & 1  & 2  \\
& ArFi LE-3   & 4   & 9   & 1  & 10  & 3 & 1  & 3 \\
& ArFi LE-4   & 5  & 12  & 1  & 13  & 3  & 1  & 4  \\
& ArFi SADP   & 3 & 3  & 1 & 5 & 5 & 3  & 1 \\
& ArFi SAQP   & 3  & 2  & 1  & 7 & 7 & 10 & 1  \\
% & SAB  & \multicolumn{1}{l}{}  & \multicolumn{1}{l}{} & \multicolumn{1}{l}{} & \multicolumn{1}{l}{}  & \multicolumn{1}{l}{}  & \multicolumn{1}{l|}{}  & 2 \\
\hline
\multicolumn{1}{c}{\multirow{2}{*}{\rotatebox[origin=c]{90}{\textbf{EUV}}}} & LE  & 1 & 3  & 1  & 3   & 3  & 0 & 1  \\
\multicolumn{1}{c}{} & SA LE-2 & 5 & 6 & 1 & 8 & 7 & 3 & 2 \\ \bottomrule
\end{tabular} 
}
\vspace{-2em}
\end{table}

%% file: tables/tab_parameters.tex
\begin{table}[t]
\caption{\name~input modeling parameters}
\vspace{-1.5em}
\label{tab:parameters}
\begin{center}
\scalebox{0.85}{
\begin{tabular}{cll} \toprule
\textbf{Parameter}  & \textbf{Description} &\textbf{Range}\\ \hline \hline
$\textrm{M}$ & Number of BEOL metal layers & From HW design (M1 -- M20)\\ 
$\textrm{\textrm{p}}$ & Process technology node & 180 nm -- 3 nm \\
$\textrm{c}$ & Patterning complexity & From fab (DUV, EUV, SADP, etc.) \\
\textrm{f} & Process flow & From fab \\
$\textrm{Y}$ & Fab yield & From fab (0 -- 1) \\ 
$\textrm{A}$ & Die area & From HW design (cm$^2$) \\

% $\textrm{N}_\textrm{r}$ & Number of ICs & From HW design \\ 
% $\textrm{K}_\textrm{r}$ & IC packaging footprint & 0.15 kg CO$_2$ \\ 
% $\textrm{A}$ & IC Area & From HW design (cm$^2$) \\ 
% $\textrm{MPA}$ & Procure materials & $\sim$0.50kg \textrm{CO$_2$} per cm$^2$ \\ 
% $\textrm{EPA}$ & Fab energy & 0.8-3.5 kWh per cm$^2$ \\ 
% $\textrm{CI}_{\textrm{use}}$ & HW CO$_2$ intensity & 30-700 g CO$_2$ per kWh \\ 
% $\textrm{CI}_{\textrm{fab}}$ & Fab CO$_2$ intensity & 30-700 g CO$_2$ per kWh \\ 
\bottomrule
\end{tabular}
}
\end{center}
\vspace{-2.6em}
\end{table}

%% file: tables/tab_ASAP7_masks.tex
\begin{table}[t!]
\centering
\caption{Approximate lithography process steps and masks per metal line process for ASAP7 PDK FEOL, MOL \& BEOL}
\label{tab:ASAP7_num_masks}
\vspace{-1em}
\scalebox{0.8}{
\begin{tabular}{cllllccc}
\toprule
\multicolumn{2}{c}{\textbf{Layers}} & \multicolumn{1}{l}{\textbf{M$_\text{pitch}$}} & \textbf{Metal} & \textbf{Via}~\cite{asap72017}   & \textbf{\# Litho steps} & \textbf{E$_\text{litho}$} & \multicolumn{1}{l}{\textbf{\# PFAS$_\text{litho}$}} \\ 
\hline \hline
\multirow{4}{*}{\rotatebox[origin=c]{90}{\textbf{FEOL}}}   & Fin       & 27   & SAQP  & -  & 2     & 1   & 1   \\
& Active    & 108   & EUV LE   & - & 3  & 10  & 1   \\
& Gate      & 54    & SADP  & - & 3   & 1  & 1  \\
& SDT  & 54  & EUV LE   & -  & 3   & 10 & 1  \\
\hline
\multirow{3}{*}{\rotatebox[origin=c]{90}{\textbf{MOL}}}    & LISD  & 54  & EUV LE   & -  & 3   & 10 & 1  \\
& LIG       & 54  & EUV LE & -    & 3   & 10    & 1  \\
& VIA0      & 25  & -  & EUV LE & 3  & 10   & 1  \\
\hline
\multirow{9}{*}{\rotatebox[origin=c]{90}{\textbf{BEOL}}}  & M1  & 36   & EUV LE  & EUV LE  & 6  & 20  & 2  \\
& M2        & 36   & EUV LE   & EUV LE   & 6  & 20  & 2   \\
& M3        & 36   & EUV LE   & EUV LE  & 6  & 20   & 2   \\
& M4        & 48   & SADP     & ArFi LE-2  & 9   & 3 & 3    \\
& M5        & 48   & SADP     & ArFi LE-2  & 9  & 3    & 3    \\
& M6        & 64    & SADP    & ArFi LE-2  & 9   & 3  & 3   \\
& M7        & 64    & SADP    & ArFi LE-2  & 9   & 3  & 3   \\
& M8        & 80    & ArFi LE & ArFi LE  & 6  & 2  & 2   \\
& M9        & 80  & ArFi LE  & ArFi LE  & 6  & 2  & 2  \\
% \hline
% \multicolumn{1}{l}{}  &   & \multicolumn{1}{l}{} &   & \textbf{TOTAL} & 86  & 128 & 29 \\
\bottomrule
\end{tabular}
}
\vspace{-1.6em}
\end{table}

% \begin{tabular}{clllccc}
% \textbf{Layer} & \textbf{Pitch}~\cite{asap72017} & \textbf{Metal} & \textbf{Via} & \multicolumn{1}{c}{\textbf{\# Litho steps}} & \multicolumn{1}{c}{\textbf{E$_{\text{litho}}$}} & \multicolumn{1}{c}{\textbf{\# Masks}} \\ \hline \hline
% M1          & 36nm        & EUV LE    & EUV LE    & 6  & 20  & 2 \\
% M2          & 36nm        & EUV LE    & EUV LE    & 6  & 20  & 2 \\
% M3          & 36nm        & EUV LE    & EUV LE    & 6  & 20  & 2 \\
% M4          & 48nm        & ArFi  LE-2 & ArFi  LE-2 & 12 & 4   & 4 \\
% M5          & 48nm        & ArFi  LE-2 & ArFi  LE-2 & 12 & 4   & 4 \\
% M6          & 64nm        & ArFi  LE-2 & ArFi  LE  & 9  & 3   & 3 \\
% M7          & 64nm        & ArFi  LE-2 & ArFi  LE  & 9  & 3   & 3 \\
% M8          & 80nm        & ArFi  LE  & ArFi  LE  & 6  & 2   & 2 \\
% M9          & 80nm        & ArFi  LE  & ArFi  LE  & 6  & 2   & 2 \\
% \hline
% & & & \textbf{TOTAL}   &  \multicolumn{1}{c}{72}  &  \multicolumn{1}{c}{78}  &  \multicolumn{1}{c}{24}   \\
% \hline
% \end{tabular}
% }
% \vspace{-1.5em}
% \end{table}

%% file: text/methodology.tex
% \begin{figure}[t!]
% \centering
% \includegraphics[width=0.8\columnwidth]{fig/PFAS_DATE2025_validation_v1.pdf}
% \vspace{-0.5cm}
% \caption{Evaluation of our PFAS analytical model versus TechInsights PFAS-containing chemicals volume trends. Our model trends closely follow the PFAS volume measurements reported by TechInsights. We leverage process complexities, fabrication process flows, and metal stacks to provide sufficient flexibility and versatility to quantify PFAS in ICs.}
% \vspace{-1.8em}
% \label{fig:PFAS-validation}
% % \vspace{-1.1em}
% \end{figure}

\vspace{-0.5em}
\section{Evaluation Methodology}
\label{sec:methodology}
\vspace{-0.3em}
% \begin{figure}[t!]
% \centering
% \includegraphics[width=0.8\columnwidth]{fig/PFAS_DATE2025_validation_v1.pdf}
% \vspace{-0.5cm}
% \caption{Evaluation of our PFAS analytical model versus TechInsights PFAS-containing chemicals volume trends. Our model trends closely follow the PFAS volume measurements reported by TechInsights. We leverage process complexities, fabrication process flows, and metal stacks to provide sufficient flexibility and versatility to quantify PFAS in ICs.}
% \vspace{-1.8em}
% \label{fig:PFAS-validation}
% % \vspace{-1.1em}
% \end{figure}

\begin{figure}[t!]
\centering
\includegraphics[width=0.8\columnwidth]{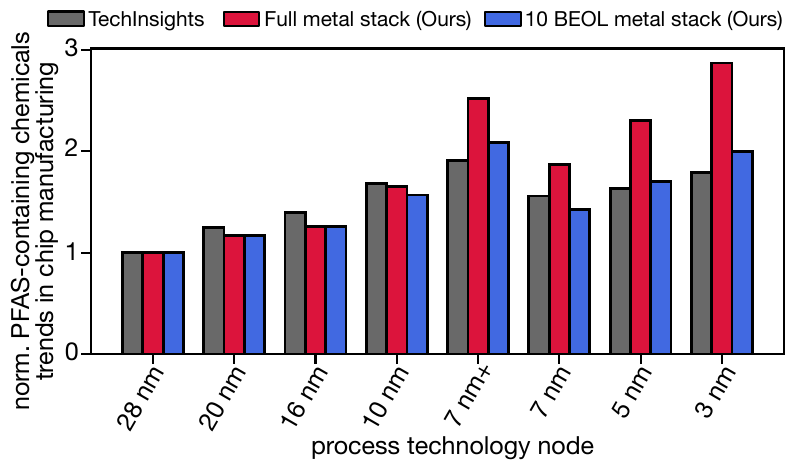}
\vspace{-0.4cm}
\caption{Evaluation of our PFAS analytical model versus TechInsights PFAS-containing chemicals volume trends. Our model trends closely follow the PFAS volume measurements reported by TechInsights. We leverage process complexities, fabrication process flows, and metal stacks to provide sufficient flexibility and versatility to quantify PFAS in ICs.}
\vspace{-1.6em}
\label{fig:PFAS-validation}
% \vspace{-1.1em}
\end{figure}

\textbf{PFAS Quantification Validation.} We validate our modeling methodology against the volume of PFAS-containing chemicals per technology node modeled by TechInsights~\cite{TechInsights2024PFASblog}. In Figure~\ref{fig:PFAS-validation}, we show the PFAS trends in manufacturing across process technology node normalized to 28 nm. Our model exhibits similar trends to the volume of PFAS-containing chemicals quantified by TechInsights. However, TechInsights provide only coarse-grained data of PFAS and only post 28 nm technology. Our PFAS model is a parametric and predictive model that builds on foundries process flows, process complexities, and metal stacks to provide sufficient flexibility and versatility to quantify and reduce PFAS use in chip manufacturing at the design phase.

% Using our \name~we exhibit similar trends as the amount of PFAS modeled from 28 nm to 3 nm process technology node (Figure~\ref{fig:PFAS-validation}). 

\textbf{Carbon Footprint.} To quantify the carbon footprint of ICs we integrate, ACT~\cite{ACTGupta2022}, an architectural carbon modeling tool into our framework for holistic sustainable computing design.

\textbf{Power-Performance-Area (PPA) Analysis.} We perform synthesis and place-and-route of a systolic array and ARM Cortex-M0~\cite{arm-cortex-m0} on Cadence Genus™ and Innovus™ using the academic ASAP7 PDK~\cite{asap72017}. 

%% file: text/results.tex
\vspace{-0.3em}
\section{Case Studies for PFAS-Aware Computing Design}
\label{sec:results}
% Especially due to the current lack of PFAS-free alternatives in photolithography and integrated circuits manufacturing. 
% Researchers and designers across the computing stack have an opportunity to identify trade-offs and incorporate optimizations for lower environmental impacts of PFAS at the design phase. 
% In this section, we present case studies for PFAS-aware computing system design. 
In this section, we first show there is contention between embodied carbon versus PFAS as advanced technology nodes adopt EUV lithography. Second, we quantify the PPA, embodied carbon, and PFAS trade-offs of manufacturing a systolic array~\cite{systolic1985} with different number of BEOL (i.e. routing) metal layers. Third, we illustrate the overall PFAS benefit for optimizing a system on chip (SoC) with less BEOL metal layers versus area trade-offs.
% using a DNN training accelerator.
% methods to optimize designs with respect to total carbon, PFAS, and power-performance-area constraints. 
% ~\cite{arm-cortex-m0}

% \begin{figure}[t!]
% \centering
% \includegraphics[width=\columnwidth]{fig/PFAS_TSMC-perf-power-PFAS-emb_v4.pdf}
% \vspace{-0.8cm}
% \caption{Normalized performance, power~\cite{TSMCn3, TSMCn7}, PFAS-containing chemicals, and embodied carbon for 1 cm$^{2}$ chip versus process technology node. We show the range of embodied carbon based on the carbon intensity, ranging from 100\% renewable energy to coal-powered, of a semiconductor fabrication facility. Manufacturing an IC at a 7 nm technology node using EUV uses 20\% less volume of PFAS-containing chemicals, compared to a 7 nm  node manufactured with DUV immersion 193 nm, while resulting in better chip power and performance.}
% \vspace{-1.3em}
% \label{fig:PFAS-TSMC-scaling}
% % \vspace{-1.1em}
% \end{figure}

% \begin{figure}[t!]
% \centering
% % \vspace{-0.45em}
% \includegraphics[width=\columnwidth]{fig/PFAS_DATE2025_systolic_plot_v3.pdf}
% \vspace{-2.5em}
% \caption{We show (a) place-and-route layout of a tiled 6$\times$6 systolic array and a single MAC unit in ASAP7; (b) the power, delay, area, embodied carbon, and number of PFAS layers trade-offs of systolic array designed to Metal 7 versus Metal 3. Reducing the number of BEOL metal stack from M7 to M3 results in  an overall 1.7$\times$ less PFAS across FEOL, MOL, and BEOL with negligible power, area, and delay penalties.}
% \vspace{-1.8em}
% \label{fig:systolic-plot}
% \end{figure}

\vspace{-0.3em}
\subsection{PFAS-Power-Performance-Carbon Design Trade-offs}
\label{sec:PFAScarbon}
\begin{figure}[t!]
\centering
\includegraphics[width=\columnwidth]{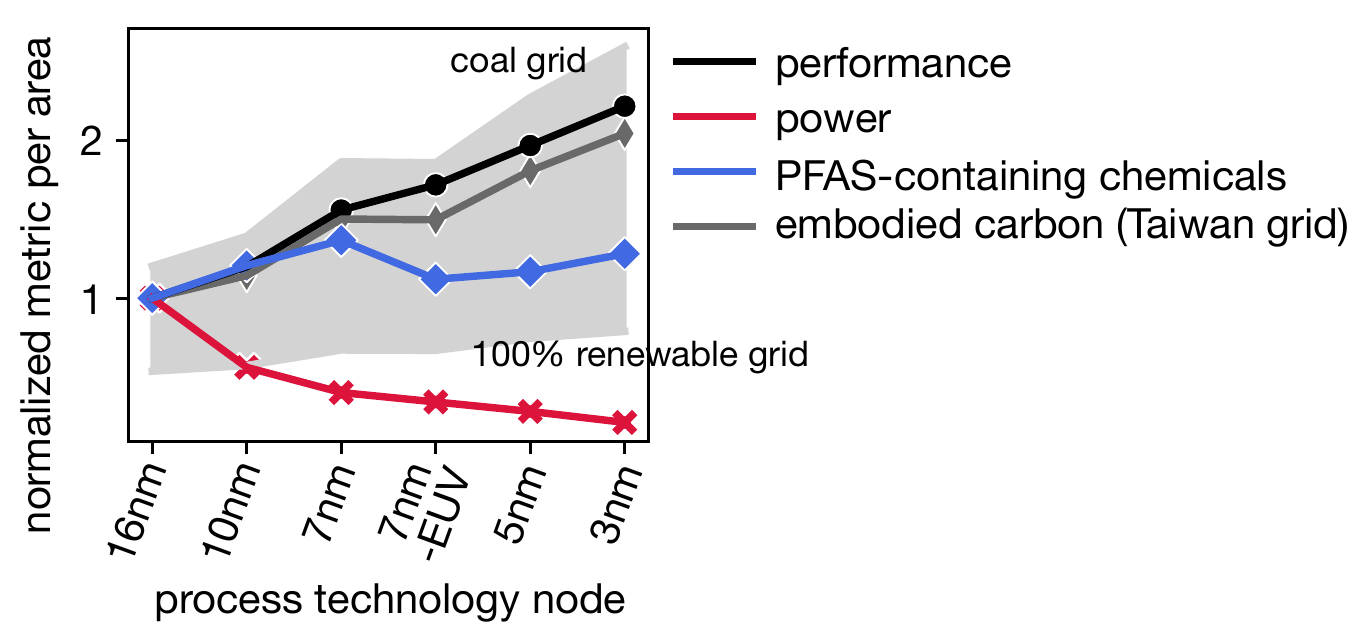}
\vspace{-0.8cm}
\caption{Normalized performance, power~\cite{TSMCn3, TSMCn7}, PFAS-containing chemicals, and embodied carbon for 1 cm$^{2}$ chip versus process technology node. We show the range of embodied carbon based on carbon intensity, ranging from 100\% renewable energy to coal-powered, of a semiconductor fabrication facility. Manufacturing an IC at a 7 nm technology node using EUV uses 18\% less PFAS-containing layers, compared to a 7 nm  node manufactured with DUV, while resulting in better power and performance.}
\vspace{-1.2em}
\label{fig:PFAS-TSMC-scaling}
% \vspace{-1.1em}
\end{figure}

\vspace{-0.2em}
To truly design more sustainable computing systems, designers need to account for environmental impacts, such as carbon footprint and PFAS, along with the conventional metrics of performance and power. In Figure~\ref{fig:PFAS-TSMC-scaling}, we show the performance, power, PFAS, and embodied carbon trends for 1 cm$^{2}$ chips from 16 nm to 3 nm (x-axis) process technology nodes based on TSMC's scaling~\cite{TSMCn7, TSMCn3}. While performance and power improve with scaling, the environmental impacts of manufacturing, including embodied carbon and PFAS, do not necessarily improve with more advanced technology nodes. For embodied carbon, the general trend is increasing with more advanced technology nodes, due to increasing energy consumption with patterning complexity~\cite{imec2020}. For PFAS-containing layers, the 7 nm (EUV) and 5 nm process technology nodes can incur lower environmental impact than 7 nm (DUV) and even 10 nm, depending on the BEOL metal stack (refer Figure~\ref{fig:PFAS-validation}), design area savings, and yield. 
% , systems may have different PFAS trade-offs

\begin{figure}[t!]
\centering
% \vspace{-0.45em}
\includegraphics[width=\columnwidth]{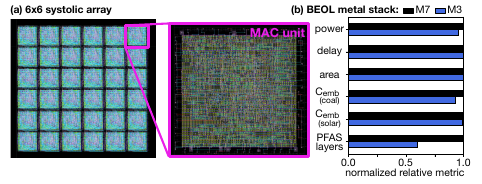}
\vspace{-2.5em}
\caption{We show (a) place-and-route layout of a tiled 6$\times$6 systolic array and a single MAC unit in ASAP7 PDK; (b) the normalized power, delay, area, embodied carbon, and PFAS layers of systolic array designed up to Metal 7 versus Metal 3. Reducing the number of BEOL metal stack from M7 to M3 results in an overall 1.7$\times$ less PFAS across FEOL, MOL, and BEOL with negligible power, delay, and area penalties.}
\vspace{-1.6em}
\label{fig:systolic-plot}
\end{figure}

To illustrate, manufacturing a 7 nm process technology node design with EUV as described in~\cite{imec2020,TSMCn7}, is more environmentally sustainable than using DUV immersion multiple patterning~\cite{imec2020} for the same metal layer stack, resulting in both lower embodied carbon and 18\% less PFAS-containing layers (20\% less volume of PFAS-containing chemicals~\cite{TechInsights2024PFASblog}) during manufacturing. This is because direct EUV helps reduce the number of masks and manufacturing steps, including lithography, deposition and etching. One direct EUV mask can replace up to five immersion DUV masks, with better patterning yield and shorter fabrication time. EUV consumes on average 10$\times$ more power per tool than traditional DUV, but replaces multiple processing steps in fabrication~\cite{imec2020}. 
% The total number of steps decreases from 1137 steps to 925 steps across First-end-of-line (FEOL), Middle-of-line (MOL), and Back-end-of-line (BEOL)~\cite{imec2020}.

%%

% \begin{figure}[t!]
% \centering
% \includegraphics[width=\columnwidth]{fig/PFAS_DATE2025_MAC_metal_screenshot_v2.pdf}
% \vspace{-2em}
% \caption{Place-and-route layouts of systolic array MAC units routed unconstrained to Metal 7 (left) and constrained to use less BEOL metal layers down to Metal 3 (right) using ASAP7 PDK.}
% \vspace{-1.4em}
% \label{fig:MAC-metal-screenshot}
% % \vspace{-1.1em}
% \end{figure}
\vspace{-0.5em}
\subsection{BEOL Metal Layers Reduction for PFAS Optimization}
\label{sec:metalstack}
\vspace{-0.2em}
%% Cortex M0 metal layers case study (G1 first project)
% In Figure~\ref{fig:PFAS-metal_layers} we quantify the performance versus PFAS versus total carbon trade-offs of manufacturing a systolic array with different number of BEOL (i.e. routing) metal layers. 
%Cortex-M0~\cite{arm-cortex-m0}
% Intuitively and similar to embodied carbon, the amount of PFAS used per chip is impacted by die area and yield. In addition, the number of BEOL metal layers used for routing contribute to amount of PFAS in manufacturing.
% Given amount of PFAS varies with BEOL metal layers, we quantify the trade-off between embodied carbon, PFAS, and power-performance-area, for example systolic arrays routed to different metal layers shown in Figure~\ref{fig:systolic-plot}(a). 
% \begin{figure}[t!]
% \centering
% \includegraphics[width=\columnwidth]{fig/PFAS_DATE2025_MAC_metal_screenshot_v2.pdf}
% \vspace{-2.2em}
% \caption{Place-and-route layouts of systolic array MAC units routed unconstrained to Metal 7 (left) and constrained to use less BEOL metal layers down to Metal 3 (right) using ASAP7 PDK.}
% \vspace{-1.4em}
% \label{fig:MAC-metal-screenshot}
% % \vspace{-1.1em}
% \end{figure}

\begin{figure}[t!]
\centering
\includegraphics[width=\columnwidth]{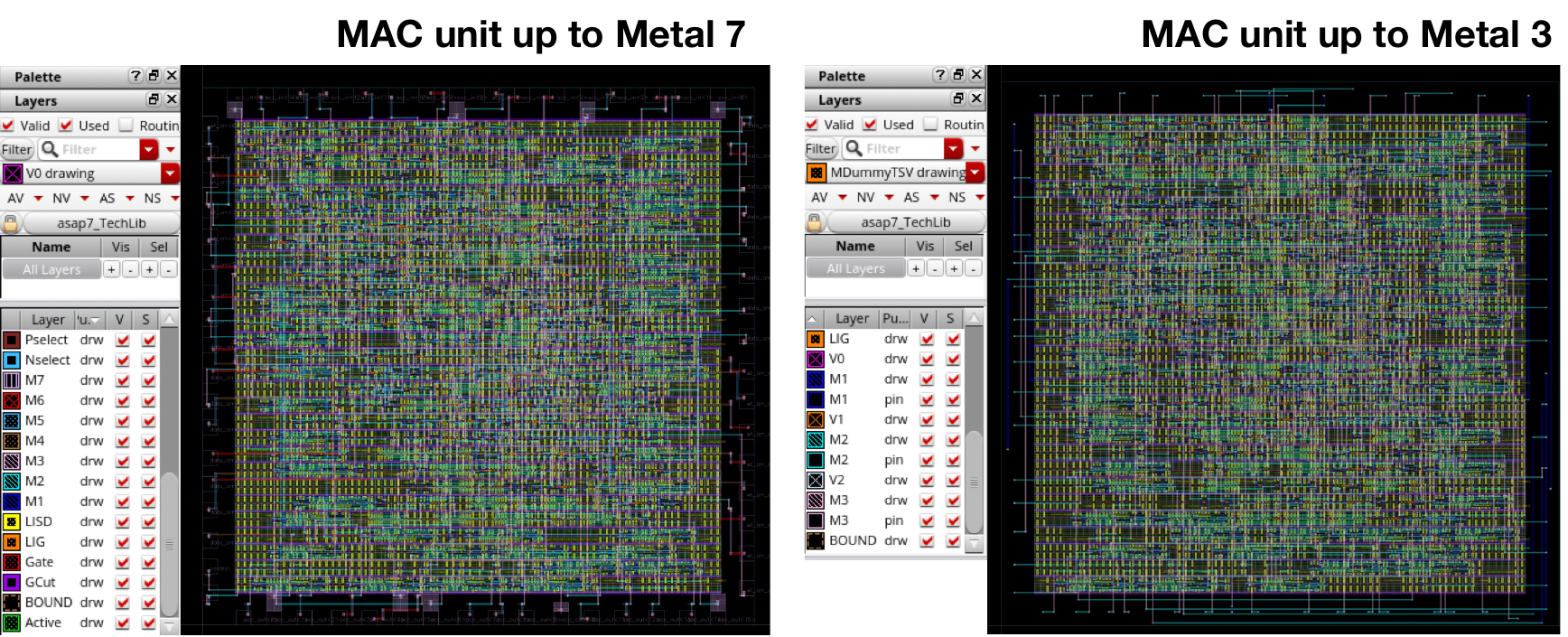}
\vspace{-2.2em}
\caption{Place-and-route layouts of systolic array MAC units routed unconstrained to Metal 7 (left) and constrained to use less BEOL metal layers down to Metal 3 (right) using ASAP7 PDK.}
\vspace{-1.2em}
\label{fig:MAC-metal-screenshot}
% \vspace{-1.1em}
\end{figure}

Given PFAS varies with number of metal layers, we quantify the trade-offs between embodied carbon, PFAS, and PPA, for example systolic arrays shown in Figure~\ref{fig:systolic-plot}(a), routed to different BEOL metal layers. 
In Figure~\ref{fig:systolic-plot}(b), we show that optimizing designs to use less BEOL metal layers can save up to 3$\times$ PFAS-containing layers across BEOL. Reducing the number of BEOL metal layers for a systolic array from Metal 7 (M7) to Metal 5 (M5) reduces the number of PFAS containing layers by 1.5$\times$, and further optimization to Metal 3 (M3) results in another 2$\times$ PFAS reduction benefit in BEOL with negligible PPA penalties due to systolic array's regularity. 
For embodied carbon footprint, reducing BEOL layers is more impactful when fabrication facilities are powered by non-renewable energy sources (e.g. coal) rather than renewables (e.g. solar), due to electricity being the primary source of embodied carbon footprint. Therefore, the primary benefit of reducing BEOL metal layers is to reduce number of PFAS-containing layers in IC manufacturing.
We show the place-and-route layouts of systolic array MAC units when unconstrained to M7 versus constrained down to M3 in Figure~\ref{fig:MAC-metal-screenshot}.

\begin{figure}[t!]
\centering
\includegraphics[width=0.83\columnwidth]{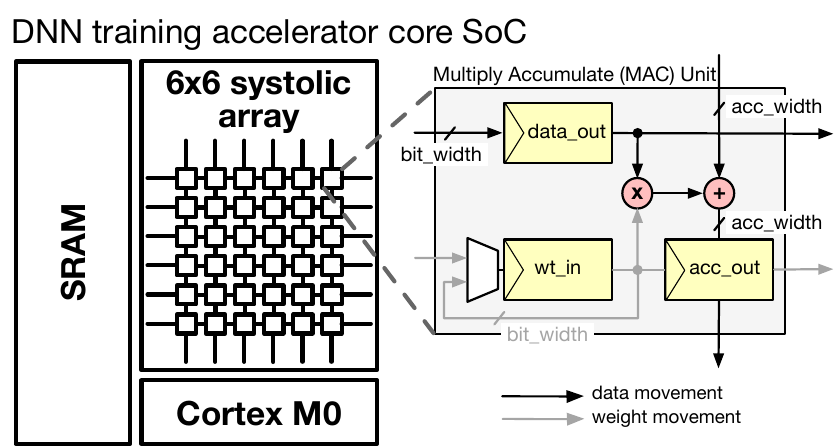}
\vspace{-0.25cm}
\caption{Overview of simplified DNN training accelerator SoC based on~\cite{CAMEL2024}.}
\vspace{-1.6em}
\label{fig:CAMEL-simplified}
% \vspace{-1.1em}
\end{figure}

\vspace{-0.3em}
\subsection{Metal Layers vs. Area Trade-off for SoC PFAS optimization}
\begin{figure}[t!]
\centering
\includegraphics[width=\columnwidth]{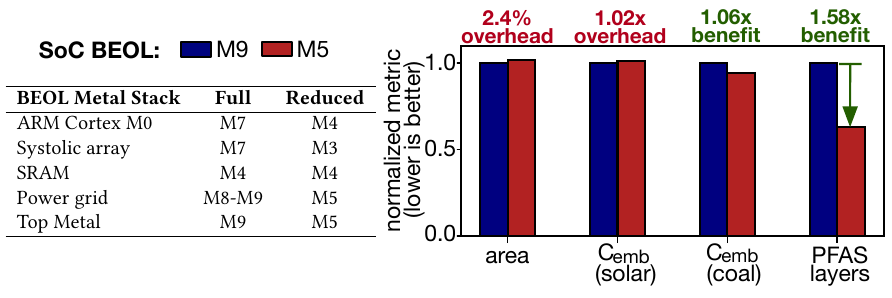}
\vspace{-2.5em}
\caption{BEOL metal layers versus area trade-off in an accelerator SoC. We optimize the SoC's BEOL metal layers from M9 to M7 and achieve a 1.58$\times$ PFAS benefit, despite a 2.4\% SoC area overhead. }
% M8 and M9 are for the power grid. (left). M5 is for the power grid. (right)
\vspace{-1.6em}
\label{fig:CAMEL-comp}
% \vspace{-1.1em}
\end{figure}

\vspace{-0.2em}
To illustrate the impact of optimizing BEOL metal layers to reduce PFAS on an SoC scale, we model a deep neural network (DNN) training accelerator~\cite{CAMEL2024}, which includes an ARM Cortex-M0, a $6\times6$ systolic array, and an on-chip SRAM, using the ASAP7 PDK metal stack (Figure~\ref{fig:CAMEL-simplified}). We do unconstrained place-and-route for the systolic array and Cortex-M0, which occupy up to M7, then optimize the BEOL metal stack routing down to M4. For the processor Cortex-M0, the area overhead increased by 1.47$\times$ to enable routing to M4. However, routing the Cortex-M0 down from M7 to M4 only results in 2.4\% increase in the total SoC area. There is no area penalty for SRAM cells since they are typically routed to M4. In Figure~\ref{fig:CAMEL-comp}, we detail the SoC's BEOL metal stack when routed up to M9 (M8-M9 for power grid) and show a 1.58$\times$ PFAS reduction benefit by optimizing to M5. 

Furthermore, we quantify the embodied carbon footprint of each SoC, based on the semiconductor fabrication facility's carbon intensity and the BEOL metal stack. In both carbon intensity cases, reducing the number of BEOL metal stack, while accounting for the area overhead, has negligible effect on chip's embodied carbon footprint. This is primarily due to the complexity of embodied carbon accounting which includes electricity consumption during fabrication (i.e. different tools and processes), materials procurement, and direct gas emissions. In contrast, the benefits of reducing number of PFAS containing layers are 10$\times$ more significant. 

% This highlights
% In Figure, we show a simplified version of the accelerator core SoC. 
% Wafer + mass production number of chips scale discussion

% \begin{wrapfigure}{rt}{0.5\columnwidth}
%   \begin{center}
%   \vspace{-2em}
%     \includegraphics[width=0.5\columnwidth]{fig/PFAS_DATE2025_SoC_v1.pdf}
%   \end{center}
%   \vspace{-1.5em}
%   \caption{M8 and M9 are for the power grid. (left). M5 is for the power grid. (right)}
%   \vspace{-5em}
%   \label{fig:CAMEL-comp}
% \end{wrapfigure}

% \subsection{PPAtC-PFAS Design Trade-offs}

% In Figure~\ref{fig:PPAtC-PFAS-opt} we optimize designs with respect to total carbon, PFAS, and power-performance-area constraints. (e.g. Pareto, uncertainty, ...)

%% file: text/opportunities.tex
\vspace{-0.3em}
\section{Opportunities \& Call to Action}
% for PFAS Optimization\\ 
%at the Design Phase
\label{sec:opportunities}
\vspace{-0.2em}
% % Researchers and designers across the computing stack have an opportunity to identify trade-offs and incorporate optimizations for lower environmental impacts of PFAS at the design phase
Optimizing the environmental impact of computing systems extends beyond carbon footprint to include materials and chemicals used in manufacturing such as PFAS.
As regulatory PFAS guidance is evolving, it is critical for the semiconductor supply chain, including IC manufacturing, to minimize and eliminate PFAS use whenever possible. As designers, we have the opportunity to identify trade-offs and incorporate optimizations in our designs to consume less PFAS during manufacturing.
In this section, we outline opportunities and future design strategies for reducing PFAS to design more environmentally sustainable computing systems.

First and foremost, more accurate and standardized PFAS quantification methods are needed as discussed in~\cref{sec:limitations}. Additionally, designers and archtiects have the opportunity to design and repurpose hardware for longer use to minimize e-waste. By extending hardware lifetime, designers can minimize the amount of e-waste sent to landfills and incineration sites, resulting in less PFAS polluting the atmosphere or leaching into the soil and aqueous streams.
 
Furthermore, there are opportunities to explore and advance the environmental sustainability of heterogeneous chiplet systems. State-of-the-art VLSI systems use multi-chip integration techniques including chiplets, i.e. assembling separately packaged die on a substrate known as interposer. Chiplets have the benefit of using small modular chips, which can help improve yield and also minimize the number of BEOL metal layers across the chip. For example, instead of routing the SoC to the maximum number of BEOL layers required (e.g. SoC to M7), each block can be manufactured up to the metal layer it requires (e.g. SRAM to M4), potentially reducing PFAS. However, chiplets also require packaging, which may introduce more PFAS and the power, performance, and carbon trade-offs versus monolithic SoCs are still unclear. This is an open area of research, which we hope \name~lays the foundation for the computing community to explore.  

% However, chiplets may also introduce more PFAS due to more packaging.
% to enable and support efforts in optimizing PFAS use in computing systems.
% \textbf{Designing hardware for longer use and repurposing hardware to minimize e-waste.} By extending the lifetime and hardware use, we minimize the amount of e-waste sent to landfills and incineration sites, resulting in less PFAS polluting the atmosphere or leaching into the soil and aqueous streams.
% \textbf{Better and standardized PFAS quantification methods.} The environmental impact of PFAS in computing is complex and understudied.
% Researchers and designers across the computing stack have an opportunity to identify trade-offs and incorporate optimizations for lower environmental impacts of PFAS at the design phase. Especially due to the current lack of PFAS-free alternatives in photolithography and integrated circuits manufacturing.
% \textbf{Opportunities for heterogenous chiplets.}
% \textbf{Designing hardware for longer use and repurposing hardware to minimize e-waste.} By extending the lifetime and hardware use, we minimize the amount of e-waste sent to landfills and incineration sites, resulting in less PFAS polluting the atmosphere or leaching into the soil and aqueous streams.
% \textbf{Better and standardized PFAS quantification methods.} The environmental impact of PFAS in computing is complex and understudied.
% We outline future steps to enable sustainable computing and reduce PFAS use and contamination from technology:

%% file: text/conclusion.tex
\vspace{-0.3em}
\section{Conclusion}
\label{sec:conclusion}
% \vspace{-0.2em}
The environmental impacts of computing systems, including carbon emissions and forever chemicals, are an escalating global concern. Semiconductor and electronics manufacturing rely heavily on the pervasive use of PFAS-containing materials, which pose uncertain bioaccumulation and human health risks. As reliance on technology grows, these hidden chemicals present urgent sustainability challenges. We propose a framework to enable designers and researchers to quantify PFAS and optimize computing systems for environmental sustainability---carbon footprint, PFAS, power, performance, and area---at the design phase. 
% Our modeling framework and datasets are open-sourced to advance sustainable computing research. 
This work aims to lay the foundation for future research and optimization efforts to reduce PFAS in computing. %the environmental impact of

% We propose, \name~to enable designers and researchers to quantify PFAS and optimize computing systems for environmental sustainability---carbon footprint, PFAS, power, performance, and area---at the design phase. 
% . As computer designers, we need to reduce the environmental impact of our computing systems, including carbon footprint and PFAS
% \textbf{Minimizing use of PFAS-containing chemicals in manufacturing.} Only a handful of technology companies, such as Apple~\cite{Apple2022whitepaper}, have announced the phasing out of PFAS in their products. More semiconductor and technology companies should invest in research and infrastructure to (1) find PFAS-free alternatives in their products, (2) safe disposal of PFAS-containing e-waste.